# Multipoint-to-point data aggregation using a single receiver and frequency-multiplexed intensity-modulated ONUs

Zichuan Zhou [1], Jinlong Wei [2], Kari A. Clark [1], Eric Sillekens [1], Callum Deakin [1], Ronit Sohanpal [1], Yuan Luo [3], Radan Slavík [4], and Zhixin Liu [1]

(1) Optical Networks Group, University College London, London, UK zhixin.liu@ucl.ac.uk
(2) Huawei Technologies Duesseldorf GmbH, Munich, Germany
(3) The Chinese University of Hong Kong (Shenzhen), Shenzhen, China
(4) Optoelectronic Research Centre, University of Southampton, Southampton, UK

**Abstract** We demonstrate 2.5-GHz-spacing frequency multiplexing capable of aggregating 64 intensity-modulated end-users using low-speed electronic and optoelectronic components. All optical network units (ONUs) achieved high per-user capacity with dedicated optical bands, enabling future large-bandwidth and low latency applications.

**Introduction**

Emerging edge cloud and latency sensitive services (e.g., 5G, tactile internet, autonomous vehicles, virtual reality) require low and stable latency connectivity between the cloud and end-users[1,2]. Such networks should be scalable to support a large number of end-users, e.g., for the internet of things[3]. Importantly, these requirements should be met using deployed legacy fibre links, which are primarily multipoint-to-point (MP2P) passive optical networks (PON), to deliver these services at low cost. In such systems, the main technical challenge is providing dedicated upstream connection from each optical network unit (ONU) with asynchronous, burst-mode upstream signals, without time or frequency domain collision[4].

The existing standardised time division multiplexing (TDM) approach, such as Ethernet PON or Gigabit PON, requires time scheduling and a large gap between consecutive optical bursts to avoid collisions at the remote splitter and consequently, cannot provide dedicated, low and stable latency services to all end-users[4]. In addition, all TDM ONUs need to operate at the full rate (e.g. 50 Gb/s for 50-GPON[5], whilst the effective per-user upstream capacity is only a factor 1/N (N is the number of ONUs[5]), resulting in a limited system capacity and inefficient ONUs.

Wavelength division multiplexing (WDM) offers a dedicated channel for each end-user and promises high per-user capacity as well as low and stable latency[4]. However, WDM significantly increases capital and operating expenses. Combined WDM-TDM approaches have lowered cost versus WDM, but the use of TDM prevents low and stable latency[6].

To provide dedicated frequency channels (and thus low latency) at low cost, multiple ONUs were multiplexed in the frequency domain (FDM) and detected by a single coherent receiver[7]. This significantly lowers capital cost[8]. Unfortunately, the demonstration of this concept only showed up to four end-users[7] and it cannot be scaled economically to support a large number of end-users, as it used expensive external cavity lasers locked to their own individual etalons.

In this work, we propose to address this by enabling low-cost ONUs locked to a frequency comb reference from an optical line terminal (OLT), representing an economical and scalable route to providing dedicated, low-latency frequency multiplexed subcarrier channels at minimal cost. We demonstrate the first 2.5-GHz-spaced frequency division multiplexed (FDM) optical access paradigm using low-cost, colour-less, intensity modulated ONUs with low-bandwidth electronics.

As shown in Fig.1a, all ONU upstream signals in our configuration are passively combined through a remote splitter and are detected simultaneously using a single coherent receiver. The FDM of ONUs is achieved by frequency locking simple ONUs' lasers to different reference comb tones distributed from the optical line terminal (OLT) (Fig.1b), resulting in 2.5-GHz-spaced optically FDM ONU upstream channels (Fig.1c). The ONUs use low-cost single-wavelength lasers of the same model which can be flexibly and stably locked to any of the 64 channels across the 160 GHz (1.2 nm in wavelength) optical bandwidth. Using <2.5 GHz electronics and subcarrier modulation intensity modulation (SCM) only, a per-user upstream data rate of up to 2.144 Gb/s using 4QAM is achieved, corresponding to an aggregated data rate of 137.2 Gb/s (net rate of 128 Gb/s).

**Experimental Set-up**

Fig. 2a shows the experimental setup. The system consisted of an OLT, a 25 km SMF-28 distributing fibre, a 1:64 remote splitting node and three branches of feeder fibres of 1 km, 4 km and 20 m length to three ONUs. The remote node was emulated using two splitters and a 9-dB-loss optical attenuator (VOA2). The total loss of the 25 km link and the remote splitting was 24 dB.

The OLT used a 30-kHz-linewidth laser emitting a 13 dBm continuous wave (CW) signal at 1550.08 nm as both the local oscillator (LO) for coherent detection of the upstream signals and the seed light for the downstream frequency comb, via a 70:30 splitter. 64 comb lines with 2.5-GHz-spacing were generated by driving a cavity-enhanced modulator with a 2.5-GHz RF signal[9]. Subsequently, the comb was amplified by an erbium doped fibre amplifier (EDFA1), scrambled by a polarisation scrambler, and attenuated before being sent to the ONUs as optical frequency references (Fig.3a).

The ONUs used the same model of single-wavelength low-cost lasers outputting 8 dBm CW[10]. The CW light was split by a 50:50 coupler and mixed with the downstream frequency comb to generate a beat note corresponding to the frequency difference between the CW and the selected reference tone for feedback frequency locking, using a proportional integral (PI) controller (Fig.2b). In this experiment, the ONU lasers were tuned to the target frequency channels using thermoelectric coolers (TECs). A 13°C temperature range is sufficient to tune the ONUs across the entire 160 GHz range. In practice, the ONUs can be tuned and controlled by the slow loop of the PI controllers for automatic channel selection. We implemented three ONUs and locked them to three neighbouring comb tones (2.5 GHz apart). For performance validation, we tuned the ONUs across the whole 160 GHz range. The upstream ONU signals were generated by modulating an electroabsorption modulator (EAM) driven with 1.072 GBaud SCM-QAM signals, generated using 4.9 GSa/s digital-to-analog converters (DACs). The digital SCM-QAM signals were generated offline using a PRBS sequence of 214-1 length, mapped to QAM symbols, shaped by a root-raise-cosine filter with a 0.01 roll-off factor, and upconverted to a carrier frequency of 0.635 GHz to generate real-value SCM-QAM signals (spectrum shown in Fig.2c). 4QAM was used for a per-user data rate of 2.1 Gbit/s. The power of the upstream signals reaching the remote splitter was -3, -5 and -6 dBm for ONU1, ONU2 and ONU3, respectively, due to different losses of the used EAMs and the feeder fibre links.

We employed dummy channels to populate the rest of the upstream channels to emulate the simultaneous transmission and detection of 64 ONUs. The dummy channels were generated by modulating a frequency comb using a Mach-Zehnder modulator (MZM) driven with 1.072 GBaud intensity-modulated SCM-QAM signals with a carrier-signal-to-power ratio of about 14 dB, which is similar to that of the ONUs' outputs. The comb for the dummy channels was tapped from EDFA1 and filtered using a waveshaper, configured to generate a flat comb with a 30 GHz bandwidth notch centred at the ONUs' frequency band. The modulated dummy channels are shown as green lines in Fig.3b.

The combined upstream signals were sent to the OLT, pre-amplified by EDFA3 (noise figure of ~5 dB), then filtered and detected by a 70 GHz bandwidth dual-polarisation coherent receiver. The waveforms were subsequently captured by a 100-GHz-bandwidth 256-GSa/s real-time oscilloscope before performing offline DSP, in which the ONU channels were digitally filtered out for demodulation. No dispersion compensation was required due to the low per-user bandwidth.

**Results and discussions**

Fig.4 shows the BER measured with regard to per-channel power for the three tested ONUs when they are locked to neighboring channels at the centre (channel 1-3, closed markers) and the edge (channel 29-31, open markers) of the optical bandwidth. Their frequency offset to centre wavelength (i.e., the LO wavelength), is $\Delta f = i \times 2.5 GHz$, where i is the channel ID. ONU1 exhibited about 4 dB higher sensitivity than ONU2&3 irrespective of modulation format because the EAM for ONU1 was optimised for 1550 nm, whilst the EAMs for the other two ONUs were optimised for 1545 nm. Considering ONU1, the sensitivities for the formats of SCM-4QAM at the hard-decision forward error correction (HD-FEC) threshold of 4.4e-3 (6.7% overhead[11]) and soft-decision threshold of 2e-2 (15.3% overhead[12]) were -44 and -47 dBm, respectively.

The BER performance across the whole 160 GHz bandwidth was characterized by tuning the three ONUs (locked to neighboring channels) from -80 GHz frequency offset (Ch -32) to 80 GHz (Ch 32). Fig.5 shows the sensitivities of ONU1 over 160 GHz for the SD-FEC and HD-FEC threshold. All 64 channels achieved sub-HD-FEC BER for 4 QAM. The average sensitivities for the HD-FEC and SD-FEC threshold for the SCM-4QAM formats were about -40 and -44 dBm, respectively. The BER degradation at the edge of the optical band was primarily due to the receiver frequency roll-off. We then calculated the aggregate upstream capacity over the whole bandwidth assuming optimised EAM (ONU1) was used, resulting in raw data rates of 137.2 Gb/s for the SCM-4QAM.

**Conclusion**

We demonstrated 2.5-GHz-spacing FDM mutlipoint-to-point data aggregation, providing a record high number of end-users with dedicated optical channels, offering low and stable latency compared to TDM. Up to a 137.2 Gb/s aggregate data rate was achieved using a single receiver and intensity-modulated colourless ONUs using low-speed, low-cost electronic and optoelectronic components, promising cost-effective scalability for MP2P connection.

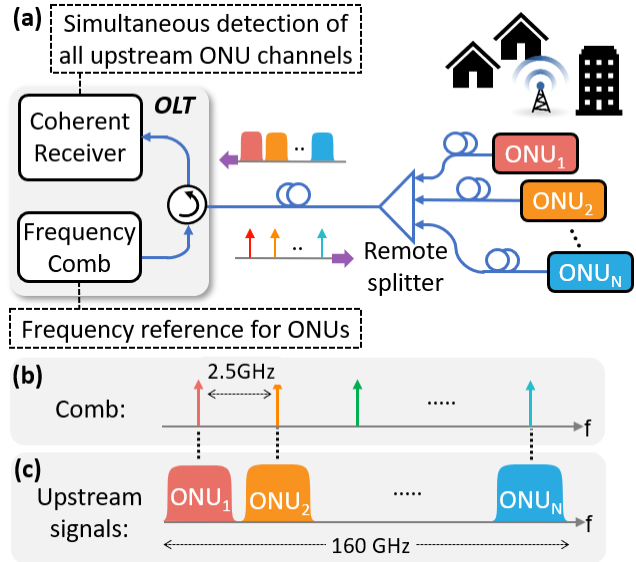

**Fig. 1**: Conceptual diagram of the proposed FDM upstream aggregation: (a) system architecture; (b) frequency comb sent to the ONUs; (c) upstream ONU signals detected by a single coherent receiver.

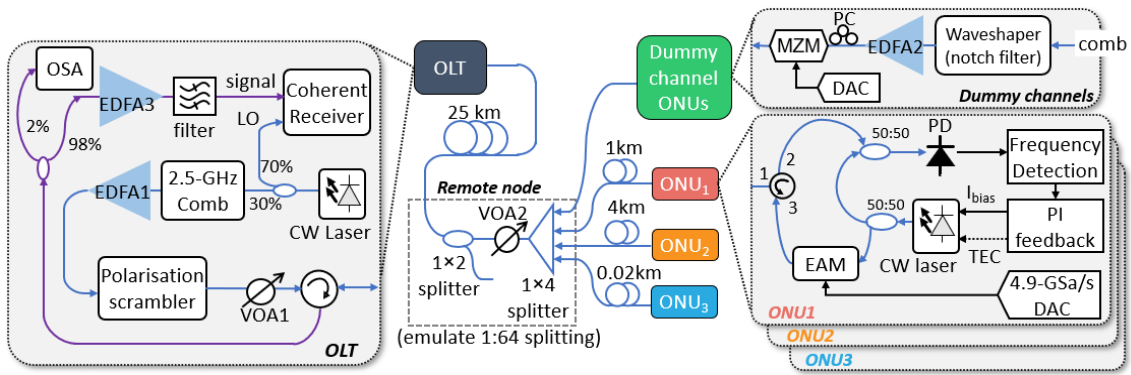

**Fig. 2**: (a) Experimental setup. Three ONUs were used in this proof-of-concept experiment; the rest of the optical bandwidth was populated using the dummy channels.

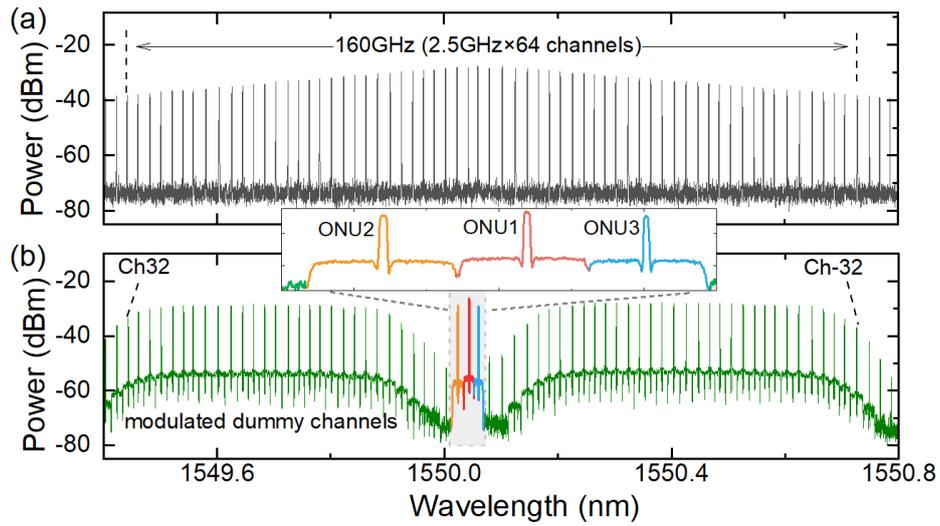

Fig. 3: Optical spectra of (a) frequency comb received at ONUs; (b) upstream ONU signals received at the OLT: red (ONU1 locked at Ch2), orange (ONU2) and blue (ONU3). Green lines show the modulated dummy channels.

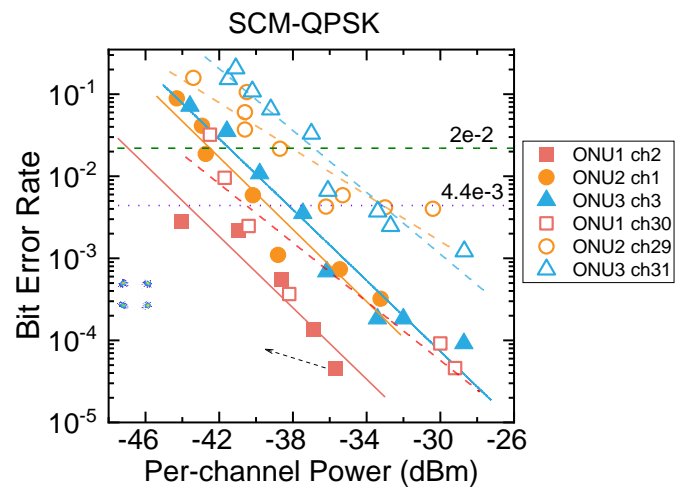

Fig. 4: BER sensitivities of the ONUs locked at the centre (solid markers, channel 1-3) and the edge (open markers, channel 29-31) of the receiver bandwidth using SCM-4QAM

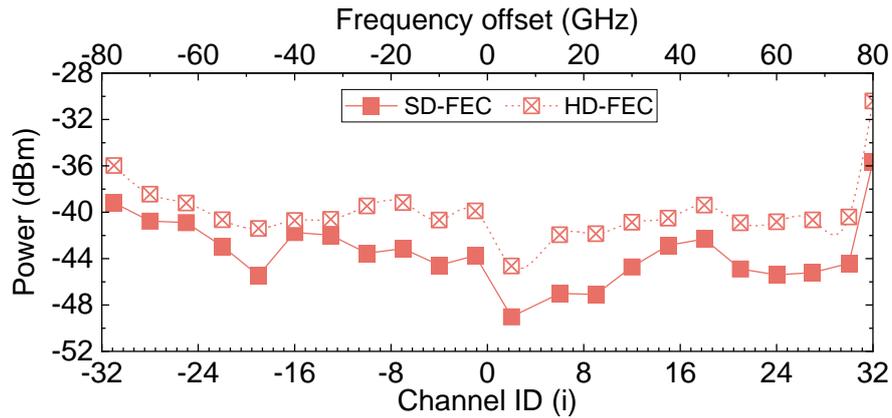

Fig. 5: Upstream power sensitivity for HD-FEC (BER of 4.4e-3) and SD-FEC threshold (BER of 2e-2). Power measured using OSA before the pre-amplifier (EDFA3) of the coherent receiver.   The relationship between the channel ID (i) and the frequency offset to centre wavelength is ∆f = i×2.5 GHz.

## Reference

[1]   "FSAN standards roadmap 2.0" 2016, Available online https://www.fsan.org/roadmap/.
[2]   N. Benzaoui, "Deterministic Latency Networks for 5G Applications", in Proc. European Conference on Optical Communications (ECOC), 2020.
[3]   N. Shibata, P. Zhu, K. Nishimura, Y. Yoshida, K. Hayashi, M. Hirota, R. Harada, K. Honda , S. Kaneko , J. Terada, and K. Kitayama, "Time Sensitive Networking for 5G NR Fronthauls and Massive Iot Traffic," J. Lightw. Technol., vol. 39, no.16, pp 5336-5343, 2021.
[4]   ITU-T G.989.2, "40-Gigabit-capable passive optical networks 2 (NG-PON2): Physical media dependent (PMD) layer specification", 2014.
[5]   D. Zhang, D. Liu, X. Wu, and D. Nesset, "Preoss of ITU-T higher speed passive optical network (50G-PON) standardization", J. Opt. Commun. Netw., vol. 12, no. 10, pp. D11-D108, Oct. 2020.
[6]   K. Grobe, and J. P. Elbers, "PON in adolescence: from TDMA to WDM-PON," IEEE Commun. Magzine, vol. 46, no. 1, pp. 26-34, Jan. 2008.
[7]   D. Lavery, L. Galdino, Z. Liu, S. Erkılınç, and P. Bayvel,"A 32 × 10 Gb/s OLT using a single ultra-wide bandwidth dual local oscillator coherent receiver," in IEEE Photonics Conf. (IPC) Part II, Orlando, Florida, 2017.
[8]   J. Bäck, P. Wright, J. Ambrose, A. Chase, M. Jary, F. Masoud, N. Sugden, G. Wardrop, A. Napoli, J. Pedro, M. Iqbal, A. Lord, D. Welch, "CAPEX Savings Enabled by Point-to-Multipoint Coherent Pluggable Optics Using Digital Subcarrier Multiplexing in Metro Aggregation Networks," 2020 European Conference on Optical Communications (ECOC), Brussels, Belgium, 2020, pp. 1-4
[9]   S. Xiao, et al., "Toward a low-jitter 10 GHz pulsed source with an optical frequency comb generator," Opt. Express, vol. 16, no. 12, 2008.
[10] J. O'Carroll, R. Phelan, B. Kelly, D. Byrne, L. P. Barry, and J. O'Gorman, "Wide temperature

range 0 < T < 85 °C narrow linewidth discrete mode laser diodes for coherent communications applications," Opt. Express, vol. 19, no. 26, pp. B90-B95, 2011.

[11] Y. Miyata, K. Kubo, K. Onohara, W. Matsumoto, H. Yoshida, and T. Mizuochi, "UEP-BCH product code based hard-decision FEC for 100 Gb/s optical transport networks," in Proc. Opt. Fiber Commun. Conf. (OFC), 2012

[12] OpenZR+ Specifications, v1.0, 2020.